\documentclass[aps,prl,reprint,groupedaddress]{revtex4-2}

\usepackage{graphicx}
\usepackage{dcolumn}
\usepackage{bm}

\usepackage{upgreek} 

\newcommand{\ltsim}{\protect\raisebox{-0.5ex}{$\:\stackrel{\textstyle <}{\sim}\:$}}
\newcommand{\gtsim}{\protect\raisebox{-0.5ex}{$\:\stackrel{\textstyle >}{\sim}\:$}}

\begin{document}


\title{Study of curling mechanism by precision kinematic measurements of curling stone's motion}


\author{Jiro Murata}
\email[]{jiro@rikkyo.ac.jp}
\affiliation{Department of Physics, Rikkyo University, Tokyo 171-8501, Japan}


\date{\today}

\begin{abstract}
Why do curling stones curl? That is a question physicists are often asked, yet no answer has been established.
Stones rotating clockwise curl right, contrary to our naive expectations. After a century of debate between contradicting hypotheses, this paper provides the answer based on experimental evidence.
A digital image analysis technique was used to perform precision kinematic measurements of a curling stone's motion to identify the curling mechanism. 
We observed a significant left-right asymmetric friction due to velocity dependence on the friction constant. 
Combined with the discrete point-like nature of the friction between ice and stone, swinging around slow-side friction points has been concluded as the dominant origin of the curling. 
Many new angular momentum transfer phenomena have been found, supporting this conclusion. 
\end{abstract}


\maketitle

\section*{Introduction}
\begin{figure}[ht]
\centering
\includegraphics[width=8.6cm]{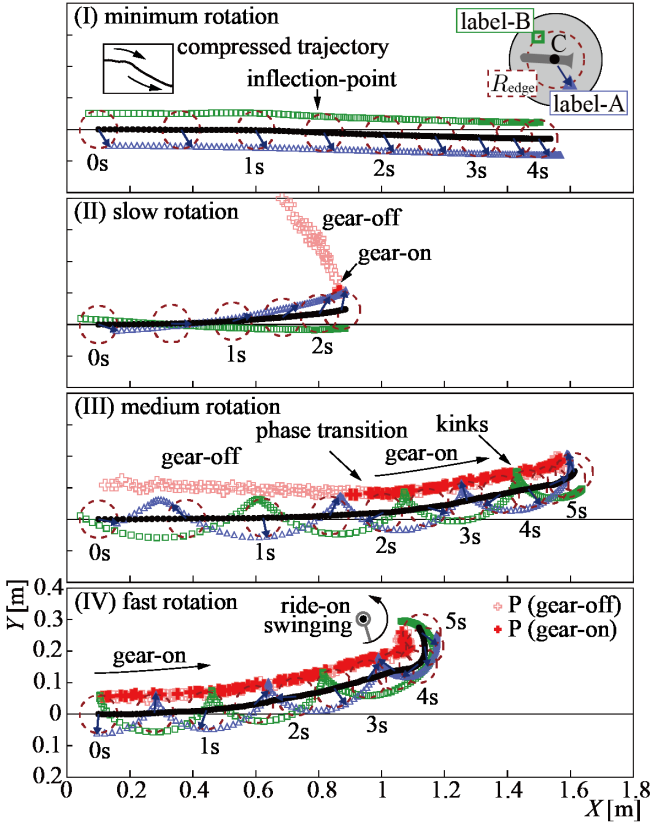}
\caption{\label{XY} 
Examples of the curling stone's trajectories for labels A, B, and C in the horizontal plane plotted for every $\Delta t$ (but $15\Delta t$ for the dotted circles and arrows).
See Fig.\ref{normanclature} for the definitions.
Raw A, B positions at $R=78\,{\rm mm}$ were corrected to be placed at $R_{\rm edge}=60\,{\rm mm}$.
We obtained data sets with initial conditions of (I)  $(|\omega_0|<0.3\,{\rm rad/s})$ for 18 low-speed $( v_0=0.3\text{--}0.7\,{\rm m/s})$ and 20 high-speed $(0.7\text{--}1.2 \,{\rm m/s})$ shots, (II) $(0.6\text{--}1.5\,{\rm rad/s})$ for 19 low-speed shots, (III) $(2\text{--}5\,{\rm rad/s})$ for 47 low-speed shots, and (IV) $(6\text{--}9\,{\rm rad/s})$ 18 low-speed shots.  
Gear-on(off) : $R_{\rm rot}\leq(>) \,65\,{\rm mm}$.
}
\end{figure}

As one of the Winter Olympics events, the curling competition is attracting more and more attention.
Along with the fun of the sport, there has been a lot of discussion about why the curling stone's trajectory bends, i.e., curls, just like the question of the principle of a breaking ball in baseball or a lift of airplane.
The curling's mysterious behavior piques the public's interest because of its opposite direction from the naively expected curling direction, considering the friction at the front edge.
For almost a century, physicists have attempted but not succeeded in explaining the curling mechanism   \cite{Harrington1924,Harrington1930,Richardson1930,Lowndes1931,Macaulay1930,Macaulay1931,Walker1937}.
Not only that, but the situation is fraught with conflicting models, owing primarily to a lack of sufficient precise observation data.


Uniform friction over the bottom of curling stones cannot produce any systematic transverse momentum transfer. 
Therefore, possible hypotheses must include forward-backward asymmetry \cite{Macaulay1930,Macaulay1931,Walker1937,Johnston1981} or left-right asymmetry \cite{Harrington1924,Denny1998} of the friction strength. 
In addition, surface roughness is often highlighted to be necessary, which may cause discrete frictioning such as pivoting due to pebble structures on ice  \cite{Penner2001,Shegelski16PV,Shegelski2018,Tusima2011,Mancini2019} and dust and scratching on ice by the stone's rough bottom surface \cite{Kameda2020}.
If we suppose the Coulomb friction law (the dynamic friction force must be opposite to the velocity direction), the left-right asymmetry of the continuum friction cannot transfer longitudinal to the transverse momentum \cite{Walker1937}.
For this reason, many hypotheses recently proposed are based on the forward-backward asymmetry requesting stronger friction at the back edge \cite{Shegelski2000,Denny2002,Nyberg2013TL,Shegelski2016,Jensen2004,Maeno2010,Maeno2014}, or a creative idea of scratch-guide mechanism \cite{Nyberg2013,Honkanen2018,Penner2019,Penner2020,Shegelski2019}, but none of which are established.

\section*{Measurement}
A precision trajectory measurement, including the rotation degree of freedom, was performed to begin a data-driven model-independent discussion.
A digital image analysis technique, originally developed as an optical alignment system for a high-energy accelerator experiment \cite{Murata2003} and as a displacement sensor for table-top gravity experiments \cite{Murata2015,Ninomiya2017}, was used. 

The measurement was performed at Karuizawa Ice Park in Nagano.
The stone's positions were measured with a sub-millimeter resolution for each static video frame ($1920\times1080$ pixels) obtained at $29.97\,{\rm frames/s}$ using a camera set on the top view position at $1800\,{\rm mm}$ above the ice surface.
Positions of two labels A and B attached on the top surface of the stone (as Fig.\ref{XY}) were measured for each frame with a time step of $\Delta t=1/29.97 \,\rm{s}$. 
Then, positions of A and B and their center C were obtained after radial position and parallax correction as vertically projected positions on the ice plane.
The $XY$ coordinates were defined relative to the direction of the initial velocity.
$t=0$ was locally defined as the starting timing for each shot.
The resolution of the image analyzing system was $47\;\upmu{\rm m}$ for a single frame, and the systematic uncertainty including non-linearities in the calibration, was $1.1\;\rm{mm}$. 
However, the actual motion was disturbed by the messy vibration caused by the pebbles. 
All shots were made toward the $X>0$ direction, but their initial conditions were not precisely controlled because we made them manually. 
Instead, they were measured.
The stone's mass and moment of inertia were $m=18.53\pm0.04\,{\rm kg}$ and $I=0.15 \pm 0.02\,{\rm kg \,m^2}$ around the center axis, respectively.

\begin{figure*}[ht]
\includegraphics[width=16cm]{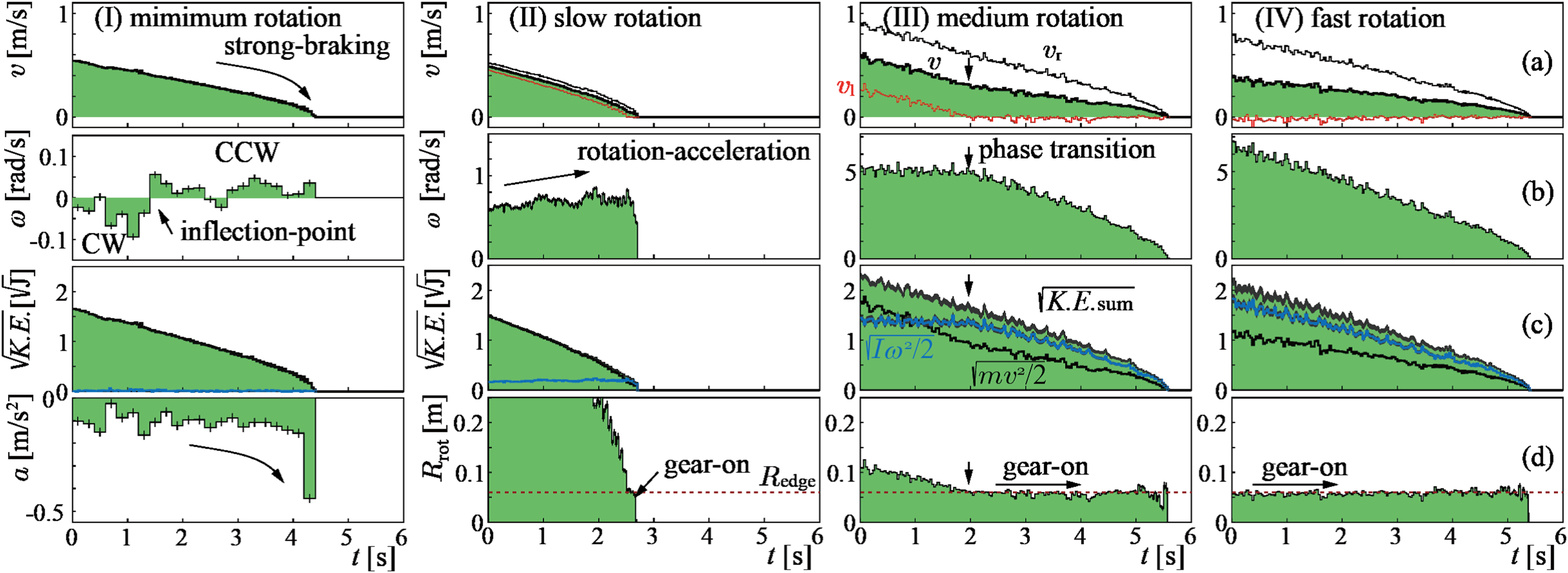}
\caption{\label{time} 
Time sequences for the same shots in Fig.\ref{XY}, of the kinematic variables; C's velocities $v(t)$, C's accelerations $a(t)$, angular velocities $\omega(t)$, kinetic energies $K.E.$, and $R_{\rm{rot}}(t)$.
The square roots of $K.E.$ were plotted for the translational components $mv^2/2$, the rotational components $I\omega^2/2$, and their sums $K.E._{\rm sum}$.
Vertical error bars were drawn if they were not negligible.
}
\end{figure*}

\section*{Results}
\begin{figure}[ht]
\centering
\includegraphics[width=6.8cm]{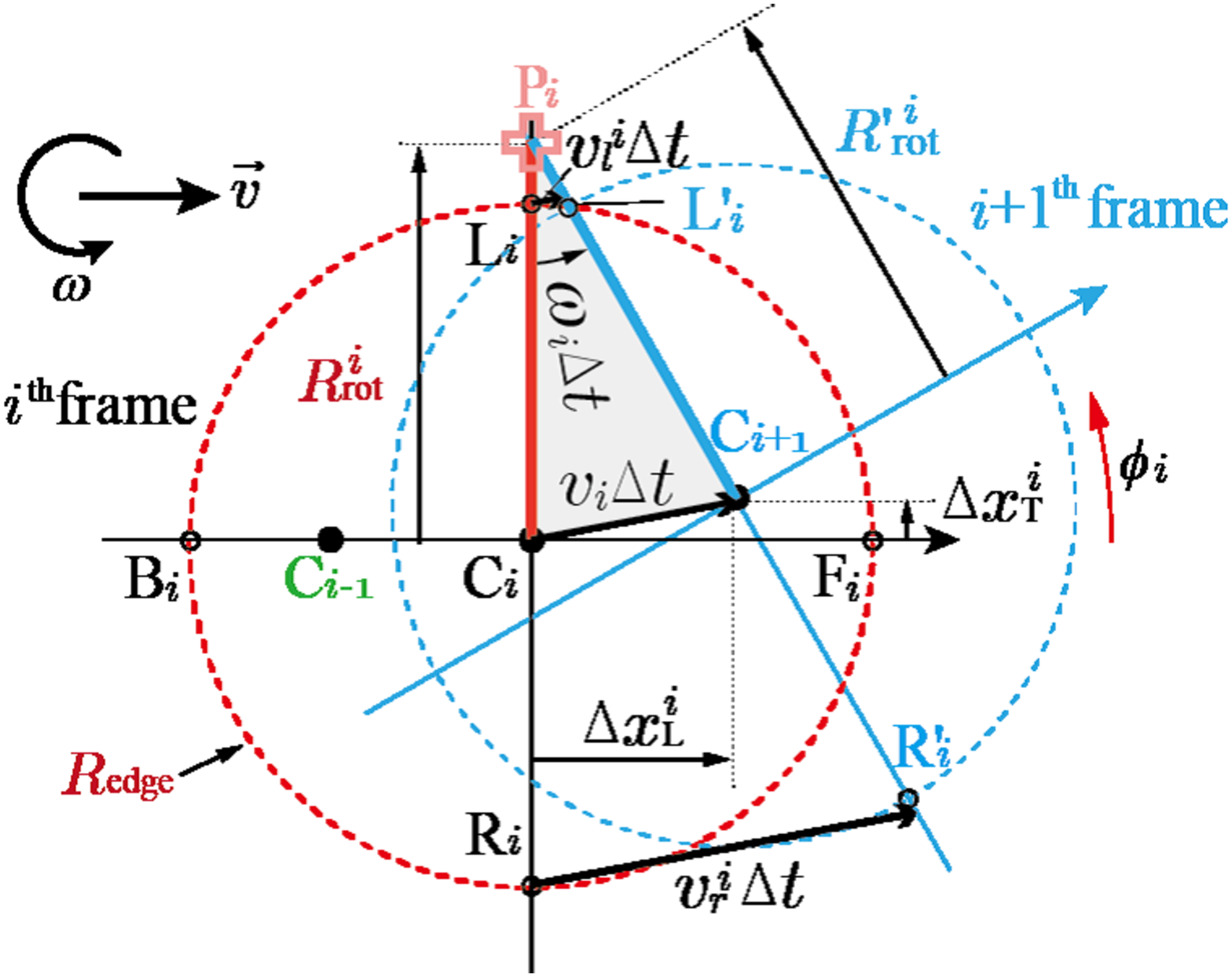}
\caption{ 
Parameter configurations. 
$i$ denotes the $i^{\rm th}$ frame at $t_i=i\Delta t$. 
The forward position ${\rm F}_i$
is determined on line
$\overline{{\rm C}_{i-1} {\rm C}_i}$,
then, $\phi_i$ is locally defined on the stone's frame in CCW from this direction.
${\rm L}_i$, ${\rm R}_i$, ${\rm F}_i$, and ${\rm B}_i$ were set on $R_{\rm{edge}}$. 
${\rm P}_i$ is defined as the intersection of lines $\overline{{\rm C}_i {\rm L}_i}$ and $\overline{{\rm C}_{i+1} {\rm L'}_i}$.
Note that ${\rm L'}_i$ is the same position fixed on the stone as ${\rm L}_i$, but that in $t=t_{i+1}$.
The ``swinging'' arm length is defined as 
$R_{\rm{rot}}^i = \overline{{\rm C}_i{\rm P}_i}=\Delta x_{\rm{L}}^i / {\rm tan} \Delta \theta_i+ \Delta x_{\rm{T}}^i$, and
${R'}_{\rm rot}^i= \overline{{\rm C}_{i+1}{\rm P}_i}=(\Delta x_{\rm{L}}^i / {\rm tan} \Delta \theta_i) / {\rm cos} \Delta \theta_i$,
where $\Delta \theta_i=\omega_i \Delta t$.
The velocities of ${\rm L}_i$ and ${\rm R}_i$ positions were approximately estimated as 
$v_{l(r)}^i=v_{i}\cdot [R_{\rm{rot}}^i-(+)R_{\rm{edge}}]/R_{\rm{rot}}^i$.
$\omega_i$ were obtained from the label's relative angular changing.
}
\label{normanclature}
\end{figure}

Fig.\ref{XY}, Fig.\ref{time}, and Fig.\ref{normanclature} show the trajectories, the time sequences of the kinematic variables for the same four typical shots, and parameter configurations. 
The differences between the $i^{\rm th}$ and $i+1^{\rm th}$ frames comprise translation and rotation.
However, they cannot be determined uniquely.
Instead, a representing point ${\rm P}$ is used.
For pure rotation around a center on the left side position, ${\rm P}$ acts as the swinging center. On the other hand, $R_{\rm rot}\equiv\overline{{\rm CP}}\rightarrow \infty$ for pure translation with no rotation.

The most symbolic phenomenon among the obtained results was the strong swinging observed before stopping, as shown in Fig.\ref{XY}-(IV) ``ride-on swinging.''
The stone swung around an almost static left position ${\rm L}\cong{\rm P}$ on the radius of $R_{\rm{edge}}$ as a simple orbital rotation.
Here $R_{\rm edge}$ was set to the inner radius of the ``running band'' (contacting ring) of the stone's bottom. 

Similar relatively strong curling was observed, as shown in Figs.\ref{XY}-(II--IV) ``gear-on'' phase.
In this phase, positions of ${\rm P}$ were not static but drifted while maintaining $R_{{\rm rot}}\cong R_{\rm{edge}}$ positions, similar to engaging gears.
In Figs.\ref{XY}-(III,IV), the velocities of ${\rm L}$ (i.e., $v_l$) were almost zero, as appeared as ``kinks'' of the label A, B's trajectories.
In fact, Figs.\ref{time}-(a,d) show $v_l\cong 0$ during $R_{\rm rot}\cong R_{\rm edge}$ gear-on phase.
As shown in Figs.\ref{XY}-(II,III), ${\rm P}$ moved from a far distance to the $R_{\rm edge}$ position during the ``gear-off'' phase and then remained there stably after this ``phase transition.''

The minimum rotation shot also revealed interesting features.
As shown in Fig.\ref{time}-(Ia), the deceleration rate of $v(t)$ was not constant, implyting the existence of velocity dependence of the friction constant $\mu$.
In fact, $a(t)$ significantly varied as a function of time, especially at the strong-braking before the end, as shown in Fig.\ref{time}-(Id).
$|\omega(t)|$ was small, but it can be highlighted that the rotation direction was significantly transitioned from CW(clockwise) to CCW (counter-clockwise) at $t\cong 1.4\,\rm{s}$ occasionally.
This transition timing coincided with the timing of the ``inflection-point'', which can be noticed if we carefully observe the compressed image of the trajectory shown in the inlet figure of Fig.\ref{XY}-(I).
It shows a transition of rightward to leftward curling.
This result implies that $\omega$ was not simply decelerating but sometimes accelerating and that the changing of $\omega$ correlated with the curling.
This ``rotation-acceleration'' phenomenon can also be found in Fig.\ref{time}-(IIb) for the slow rotation shot.

The transitioning phenomena were also found for the translational motion.
In Fig.\ref{time}-(IIIa) ``phase transition,'' the deceleration rate of $v$ suddenly decreased after $t\cong 2\,{\rm s}$.
It was at the point that the gear-on phase began.
At the same time, $\omega$ started a rapid deceleration.
This correlation can be well understood by checking the kinetic energies, $K.E.$, as shown in Fig.\ref{time}-(IIIc).
Their square roots were plotted to see quantities proportional to velocities.
The translational and rotational components exhibited the transition, but their sum did not.
This disappearance of the transition is particularly intriguing.
It means that the $K.E.$ were conserved, except for the frictional loss while transferring it between translational ($\frac{1}{2} mv^2$) and rotational ($\frac{1}{2} I\omega^2$) motions as
\begin{equation}
\label{energy}
\frac{d}{dt}(\frac{1}{2}mv^2)+\frac{d}{dt}(\frac{1}{2}I\omega^2)+{\rm frictional\;loss\;rate}=0,
\end{equation}
but the frictional loss rate was smooth without the sudden change.

The gear-off phase was also interesting, representing a situation in the actual curling games.
$\omega$ was almost constant during the gear-off phase, as shown in Figs.\ref{time}-(II,III).
The transfer of translational and rotational energies helps to explain this situation.
The rotational energy was fed by the translational energy, preventing deceleration due to rotational friction loss.
Therefore, we should not simply interpret it as minimal rotational friction constant.
In addition, the rotational energy feeds the translational energy, as shown in the deceleration relaxation of $v$ shown in Fig.\ref{time}-(IIIa). 

\begin{figure}[ht]
\centering
\includegraphics[width=8.6cm]{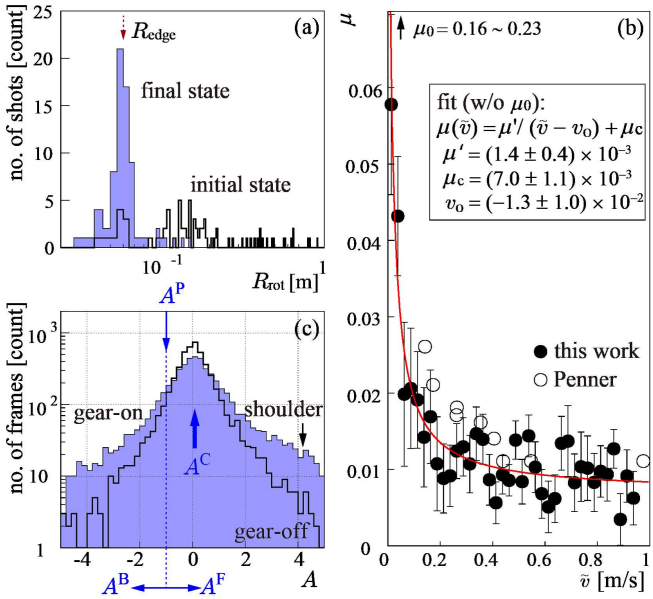}
\caption{\label{rot} (a) $R_{\rm rot}$ distribution at the initial and final states, obtained for all 73 completely stopped shots, except for minimum rotation shots,
(b) $\mu(\tilde{v})$ obtained using 38 minimum rotation shots, compared with Penner's result \cite{Penner2001},
(c) distribution of $A$ for all frames of the shots in (a).
}
\end{figure}

All shots we measured were analyzed, not only for the typical shots shown in Figs.\ref{XY} and \ref{time}.
$R_{\rm rot}$ distributions at initial and final states are plotted in Fig.\ref{rot}-(a).
It can be confirmed that the convergence $R_{\rm rot}\rightarrow R_{\rm edge}$ was the common feature of all shots, independent of the initial conditions $(v_0,\omega_0)$, except for the minimum rotation cases.
The observed peak of $R_{\rm rot}^{\rm peak}=58\,\pm\,1.3\,{\rm mm}$ for the final states was compared with the running band regions of $R_{\rm band}=65\pm 5\,{\rm mm}$.
Then, their difference should be interpreted as the azimuthal angular distribution of the actual swinging center positions, which might be spread out at approximately $\pm {\rm cos}^{-1}(R_{\rm rot}^{\rm peak}/R_{\rm band})\cong \pm27^\circ$ around $\phi=90^\circ$.

As shown in Fig.\ref{time}-(I), $\mu$ seems strongly dependent on $v$.
To confirm this, $\mu(v)$ might be estimated using the correlation between $v(t)$ and $a(t)=\mu g$ for the minimum rotation shots.
However, $\omega$ of the minimum rotation shots was not precisely zero.
Therefore, we should have considered the deceleration of $v$ caused by energy leakage from translation to rotation.
$\tilde{v}=\sqrt{2 K.E._{\rm sum}/m}$ and $\tilde{a}=d\tilde{v}/dt$ were used after rebinning the time sequence combining $8\Delta t$ to suppress statistical fluctuations.
Then, $\mu(\tilde{v})=\tilde{a}/g$ was obtained as shown in Fig.\ref{rot}-(b), using the local value of the gravitational acceleration of $g=9.796\,{\rm m/s^2}$.

The remarkable velocity dependence of $\mu$ was confirmed, showing a rapid increase before stopping.
It is crucial to understand the curling mechanism, directly implying a friction enhancement on the slower side.
Fig.\ref{rot}-(b) also shows the fitting result, which may be useful for future model calculations attempting to predict the curling trajectories as well as microscopic studies on the physics of friction.
The static friction constant $\mu_0$ measured using a spring scale was also shown, which was not used for the fitting.
Interestingly, this shape is similar to the Bragg curve, which is known as the energy loss of ions passing through matters \cite{Bragg1904,LEO}.
The analogies with the interaction of radiation with matter, known for their application in radiation therapy, are particularly interesting.

\begin{figure*}[ht]
\centering
\includegraphics[width=10.5cm]{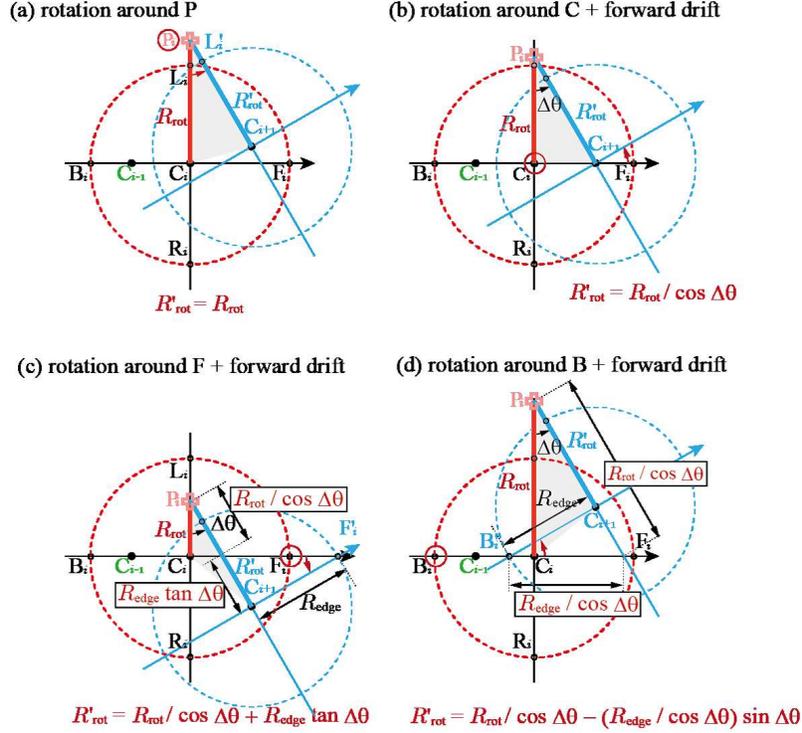}
\caption{\label{config-rot} 
Configuration for the typical rotation with forward drift cases to deduce the relationship between $R_{\rm rot}$ and ${R'}_{\rm rot}$.
}
\end{figure*}

\section*{Discussion}
Let us now attempt to understand the obtained results shown above.
First, the strong curling, as shown in Fig.\ref{XY} ``ride-on swinging,'' was a clear indication of the existence of strong point-like frictions.
It should be caused by pivoting due to relatively large pebbles on ice \cite{Shegelski2018} or dust or scratching by the rock's rough bottom \cite{Kameda2020} or their accidental coincidences. 
Therefore, these phenomena occurred by chance, with less than 50\% of our rotating shots exhibiting it.

The energy/momentum transfer between translational and rotational motions was found.
This was observed as the accelerating rotation in Figs.\ref{time}-(Ib, IIb), as the deceleration relaxation of $v$ in the gear-on phase in Fig.\ref{time}-(IIIa), and as the constant $\omega$ in Fig.\ref{time}-(IIIb).
It also meant the transfer between the orbital-angular momentum (for revolution around a fixed position) $L$ and the spin-angular momentum (for self-rotation) $S$ \cite{Farkas2003},
conserving the total angular momentum $J=L+S$ as is frequently used in quantum mechanics to treat atomic systems.
For $L=mvb$ and $S=I\omega$ ($b$ is the impact parameter, i.e., the perpendicular distance between the path of an incident particle and the center of force), the $L\leftrightarrow S$ transfer requires offset impact, i.e., $b\neq0$.
Therefore, the observed angular momentum transfer must be caused by point-like impacts at a non-zero net offset position.
Any forward-backward asymmetric friction cannot produce such angular momentum transfer because $b=0$.
Angular momentum transfer due to an offset collision to a fixed point cannot avoid swinging.
The swinging leads to the leftward curling if the impact point is at $90^\circ\ltsim \phi < 180^\circ$.
This can be confirmed as a coincidence of the curling inflection and the $\omega$ change in the minimal rotation shot as shown in Figs.\ref{XY}-(I), \ref{time}-(Ib).
It should be interesting to compare it with the production of nuclear spin polarization via offset nuclear reactions \cite{Wilczynski1973,Tanaka1986}.
In this case, the nuclei accelerated by the particle accelerator gain spin angular momentum through offset nuclear collisions.

The converging $R_{{\rm rot}}\rightarrow R_{{\rm edge}}$ can be understood as the frictional force at ${\rm L}$ being always opposite to the $v_l$ direction, suppressing $|v_l|$.
The observed converging $v_l\rightarrow 0$ also indicates that the friction is strongest at ${\rm L}$ around the $\phi$ direction in the running band.
This backward friction at ${\rm L}$ assisted rotation when $v_l>0$ via the $L\rightarrow S$ transfer, preventing deceleration of $\omega$ during the gear-off phase.
The stability of $v_l \cong 0$ and $R_{\rm rot} \cong R_{\rm edge}$ during the gear-on phase must be due to the large local static friction $\mu_0$ at ${\rm L}$.
It prevented $|v_l|$ from enlargement by sequentially switching the engaging points by next-to-next.
The frictioning points for the gear-off phase were not static but dragged while scratching the ice, which must be caused by the relatively large $\mu$ at small $v_l$.

The observed $\mu(v)$ indicated that the probability of having discrete impacts was greater at ${\rm L}$ than at ${\rm R}$
because continuum friction is not a fundamental concept but only a result of an artificial coarse-graining (averaging) treatment for many real microscopic impacts.
As a result, it should be concluded that the combination of 1. swinging around a discrete frictioning point on the ice (pivoting/scratching)
\cite{Penner2001,Shegelski16PV,Shegelski2018,Tusima2011,Mancini2019,Kameda2020} and 2. the probability of the discrete frictioning is greater at the slow-side than at the fast-side because the velocity dependence of $\mu$ \cite{Harrington1924,Penner2001} should be the dominant curling mechanism. 
The convergence $v_l\rightarrow 0$ meant the existence of force to generate strong local static frictioning $\mu\rightarrow\mu_0$ at the slow-side, which worked as the ``adhesive friction'' requested in the pivoting models \cite{Penner2001,Shegelski16PV,Shegelski2018}.

$S$ provided the left-right asymmetry of the swinging probability but was not typically the primary momentum source of the swinging.
That was $L$, which was transferred from a straight motion with the impact parameter $b$ to an orbital rotational motion with the arm length $b$, resulting in swinging for $v_l>0$, i.e., the slow rotating gear-off cases.
$S$ can directly contribute to the swinging via $S\rightarrow L$ transfer, but it was effective only for the fast rotation cases satisfying $v_l\leq 0$.
It should provide an answer to the known question \cite{Penner2001,Nyberg2013} of why the total amount of curl is not proportional to, and only weakly depends on, $\omega_0$, as shown in Fig.\ref{XY}, for the slow rotation cases.
On the other hand, $\omega_0$ dependence was becoming visible in the faster rotation cases due to the contribution from $S$ \cite{Jensen2004}.
In addition, deceleration relaxation of $v$ due to $S\rightarrow L$ transfer during the gear-on phase should be the direct answer to why extremely fast rotating stones tend to travel further \cite{Penner2001}.
The stored intrinsic rotational energy should have been used to help the translational motion overcome friction.

Finally, the forward-backward asymmetry was examined.
Although the inhomogeneous distribution of $\mu$ cannot be measured directly, we can estimate it because the discrete frictioning probability is proportional to $\mu$.
A useful tool was comparing the lengths ${R}_{\rm rot}$ and ${R'}_{\rm rot}$.
By defining
\begin{equation}
\label{Asym}
A\equiv \frac{R'_{\rm rot}-R_{\rm rot}}{R'_{\rm rot}+R_{\rm rot}} \frac{4}{\Delta \theta^2}-1,
\end{equation}
the swinging center positions were estimated.
For example,
$A^{\rm P}=-1$
for pure rotation around P without drifting, and
$A^{\rm C}=\mathcal{O}(\Delta \theta^2)$ for rotation around C with forward drifting.
For rotation around F(B) with forward driting,
$A^{\rm F(B)}\cong +(-) 2R_{\rm edge}/[R_{\rm rot}\Delta\theta]-1$.
Therefore, 
$A^{\rm F(B)}>(<)-1$.
These are because
$R'^{\rm P}_{\rm rot}=R_{\rm rot}$,
$R'^{\rm C}_{\rm rot}=R_{\rm rot}/{\rm cos}\Delta\theta$,
and
$
R'^{\rm F(B)}_{\rm rot}=R_{\rm rot}/{\rm cos}\Delta\theta+(-)R_{\rm edge} {\rm tan}\Delta\theta
$ as shown in Fig.\ref{config-rot}.

Fig.\ref{rot}-(c) shows the results, which were dominated by $A^{\rm C}$.
The realistic left side swinging with forward drifting should be distributed in $A^{\rm P} <A< A^{\rm C}$, which could not be resolved from $A^{\rm C}$ in this resolution.
As shown by ``shoulder,'' a slight asymmetry with respect to $A=0$ can be noticed, indicating that $A^{\rm F}$ was preferred to $A^{\rm B}$ for both the gear-on and -off phases.
This result indicated that $\mu({\rm front})\gtsim\mu({\rm back})$, which is naively acceptable but may lead to cause opposite curling.
Therefore, this effect should suppress the major curling, providing another possible reason for the weak dependence of the total curl on $\omega_0$.
We found no evidence for the unnaturally large asymmetry $\mu({\rm front})\ll \mu({\rm back})$ requested by the previously proposed forward-backward asymmetry models \cite{Macaulay1930,Macaulay1931,Walker1937,Johnston1981,Shegelski2000,Denny2002,Nyberg2013TL,Shegelski2016,Jensen2004,Maeno2010,Maeno2014}.

In a real curling game, the brush-sweeping is effective not only for extending the stopping range by reducing $\mu$ but also for controlling the curling.
Indeed, sweeping on the forward-right region leads to leftward curling.
It is because of the reduction of the discrete frictioning on the right side.
The unpredictable motion of the minimum rotation case is analogous to ``knuckleball'' in baseball, implying that a slight rotation should be preferred for a stable control to avoid random angular momentum transfer.
The players should also remember that the occasional ride-on swinging phenomenon randomly affects the final stopping position.

\section*{Conclusion}
In conclusion, it has been found that swinging around the discrete left-right asymmetric frictioning points is the dominant curling mechanism.
Each part of this conclusion is not perfectly new, repeatedly suggested as attractive hypotheses in previous works \cite{Harrington1924,Penner2001,Tusima2011,Shegelski16PV,Shegelski2018,Mancini2019,Kameda2020}.
Except for $\mu(v)$, most of the featured rich results supporting the conclusion presented here are new, indicating the angular momentum transfer and the point-like nature of friction.
Future model calculations must reproduce not only curling trajectories but also the phase transition and other angular momentum transfer phenomena.
This work does not propose a hypothesis but presents the principle of curling based on the model-independent experimental evidence to solve this ``mystery of the century.''



\section*{Acknowledgements}
Parts of this work were conducted as graduation research at Rikkyo University, by Ryota Aoki, Takaki Asada, Ran Ito, Daichi Sasagawa, and Ikumi Taira.
I thank Olympic athlete Tsuyoshi Yamaguchi, staff and players at Karuizawa Ice-park for their kind assistance and discussions.

\bibliography{murata}

\end{document}